\pdfoutput=1

\documentclass[11pt]{article}

\usepackage{amsfonts}
\usepackage[margin=2cm]{geometry}
\usepackage{amsmath,amssymb,amsthm,array,booktabs,cite,dsfont,graphicx,mathtools,multirow,placeins,rotating,stmaryrd,tensor,verbatim}
\usepackage[margin=2cm]{geometry}
\usepackage{slashbox}
\usepackage{caption}
\usepackage{color}
\usepackage{setspace}
\usepackage{slashed}

\usepackage[bookmarksnumbered,bookmarksopen=true,bookmarksopenlevel=1,linktocpage,pdfstartview=FitH,hyperfootnotes=false]{hyperref}
\usepackage[all]{hypcap}

\allowdisplaybreaks

\newcolumntype{M}[1]{>{$}{#1}<{$}}

\newcommand{\sst}[1]{{\scriptscriptstyle #1}}

\newcommand{\rep}[1]{\ensuremath{\mathbf{#1}}}
\newcommand\numberthis{\addtocounter{equation}{1}\tag{\theequation}}

\def\0{{\sst{(0)}}}
\def\1{{\sst{(1)}}}
\def\2{{\sst{(2)}}}
\def\3{{\sst{(3)}}}
\def\4{{\sst{(4)}}}
\def\5{{\sst{(5)}}}
\def\6{{\sst{(6)}}}
\def\7{{\sst{(7)}}}

\newcommand{\be}{\begin{equation}}
\newcommand{\ee}{\end{equation}}
\def\ba{\begin{array}}
\def\ea{\end{array}}

\newcommand{\bea}{\begin{eqnarray}}
\newcommand{\eea}{\end{eqnarray}}

\newcommand{\alg}{\mathds{A}}

\newcommand{\C}{\mathds{C}}
\newcommand{\Q}{\mathds{H}}

\newcommand{\N}{\mathcal{N}}

\begin{document}

\begin{titlepage}
\begin{center}
\hfill Imperial/TP/2014/sn/01\\
\vskip 2cm

{\Huge \bf Chiral Squaring}

\vskip 1.5cm

{\bf  S.~Nagy${}^{1,2}$}

\vskip 20pt

{\it ${}^{1}$Theoretical Physics, Blackett Laboratory, Imperial College London,\\
 London SW7 2AZ, United Kingdom}\\\vskip 5pt
 {\it ${}^2$Department of Mathematics, Instituto Superior T\'{e}cnico,\\
Av. Rovisco Pais, 1049-001 Lisbon, Portugal}\\\vskip 5pt
\texttt{snagy@math.tecnico.ulisboa.pt}

\end{center}

\vskip 2.2cm

\begin{center} {\bf ABSTRACT}\\[3ex]\end{center}
We construct the states and symmetries of $\mathcal{N}=4$ super-Yang-Mills by tensoring two $\mathcal{N}=1$ chiral multiplets and introducing two extra SUSY generators. This allows us to write the maximal $\mathcal{N}=8$ supergravity as four copies of the chiral multiplet. We extend this to higher dimensions and discuss applications to scattering amplitudes.



\vfill

\end{titlepage}

\newpage \setcounter{page}{1} \numberwithin{equation}{section} \tableofcontents 

\newpage

\section{Introduction}
The idea that supergravity is secretly a double copy of super-Yang-Mills (SYM) theories has its origins in the KLT relations of string theory \cite{kawai:1985xq}, and has found applications in the study of scattering amplitudes \cite{Bern:2008qj,Bern:2010ue,Bern:2010yg,Bianchi:2008pu,Huang:2012wr,Cachazo:2013iea,Dolan:2013isa,Chiodaroli:2014xia}, gravity anomalies from gauge anomalies \cite{Antoniadis:1992sa} and asymmetric orbifold constructions\cite{Sen:1995ff}. Recently, this correspondence has been made manifest at a more fundamental level, by giving a dictionary between gravitational and gauge fields, and by constructing the symmetries of the Lagrangian of the former from those of the latter. The global U-duality groups of supergravities built by squaring in all dimensions have been derived from the R-symmetries of the SYM theories \cite{Borsten:2013bp,Anastasiou:2013hba,Anastasiou:2015vba}. At the linearised level, the local symmetries of general covariance, $p$-form gauge invariance, local Lorentz 
invariance and local supersymmetry of the $\mathcal{N}=1$ gravitational superfield have been obtained by identifying\cite{Anastasiou:2014qba}
\be
(\mathcal{N}=1)_{SG}=(\mathcal{N}=1)_{SYM}\star (\mathcal{N}=0)_ {YM}\star \phi,
\ee
where $\star$ denotes a convolution of an $\mathcal{N}=1$ gauge superfield, a vector field and a scalar living in the biadjoint of the gauge group. This latter term can be identified as the zeroth copy in the BCJ duality \cite{Cachazo:2013iea,Hodges:2011wm,Monteiro:2014cda}.

Given the simplification in calculations brought about by this decomposition, a natural question is whether we can go further and decompose the gauge superfields in terms of chiral superfields.  Though little explored, one can find some hints of the double copy structure of gauge amplitudes in the scattering literature. Of course one can start with the double copy form of gravitational amplitudes and employ supersymmetric Ward identities to find various expressions between gluon and fermion amplitudes; however no systematic understanding of these exists. In 6 dimensions, due to the structure of the little group ($SO(4)=SO(3)\times SO(3)$) and the fact that one can build the states of the $\mathcal{N}=(1,1)$ SYM multiplet by tensoring together $\mathcal{N}=(1,0)$ and $\mathcal{N}=(0,1)$ chiral multiplets, this double copy is revealed when the amplitudes are written in the spinor helicity formalism \cite{Cheung:2009dc,Bern:2010qa}. There is no straightforward analogue in D=4 but a possible hint comes from one-
loop calculations in $\mathcal{N}=4$ SYM. It is known that the amplitude here is entirely determined by the scalar box functions. The contributions from different particles are related to the scalar contributions via supersymmetric Ward identities. To obtain the contribution from a whole multiplet we sum over all its states and those of $\mathcal{N}=4$ SYM and $\mathcal{N}=1$ chiral are related via  $\rho^{\mathcal{N}=4}=(\rho^{\mathcal{N}=1})^2$ \cite{Bidder:2005ri}.

In this paper we explore the idea of SYM multiplets themselves as a double copy. We proceed as follows. In \autoref{D=4 states}  we obtain the SUSY transformations of the $\mathcal{N}=4$ SYM multiplet from those of the N=1 chiral multiplet in four dimensions. To achieve this, we introduce extra supersymmetry generators in the chiral multiplet, obtained by a $U(1)$ rotation of the scalar states in the definition of the original $Q$'s. In a sense, this amounts to reversing the process of truncation which breaks an $\mathcal{N}=4$ gauge multiplet into an $\mathcal{N}=2$ SYM and a $\mathcal{N}=2$ hypermultiplet; however, the novelty is that  the extra SUSY generators are built entirely from the operators of the $\mathcal{N}=1$ theory.  This allows us to write the maximal supergravity in four dimensions as four copies of the (enhanced) chiral multiplet. The gauge and R-symmetries are also derived from squaring. We show that the squaring is more straightforward in $D>4$, i.e. the chiral multiplets don't need to be 
enhanced with extra Q's in \autoref{Squaring in D neq 4}. We then conclude with possible applications of our dictionary, particularly to scattering amplitudes (where the extra supersymmetry generators will become necessary when using the Ward identities) and more speculatively to off-shell superfield descriptions.

\section{D=4}\label{D=4 states}

The chiral multiplet contains a Weyl spinor and two real (or one complex) scalars, whose on-shell degrees of freedom are multiplied according to the table below:
\begin{table}[h]
\small
\begin{center}
 $\begin{array}{c|c|c|c}
 &\begin{array}{c} \tilde{\chi}^-  \end{array} &\begin{array}{c} 2 \tilde{\phi} \end{array}&\begin{array}{c} \tilde{\chi}^+  \end{array}\\
\hline
&&&\\
\begin{array}{c}\chi^- \end{array}
&\begin{array}{cccccc}  A^- \end{array} 
&\begin{array}{cccccc} 2\lambda^-\end{array} 
&\begin{array}{cccccc}   \phi\end{array}
\\
&&&\\
\begin{array}{c}2{\phi} \end{array}
&\begin{array}{cccccc} 2\lambda^- \end{array}
&\begin{array}{cccccc} 4\phi \end{array} 
&\begin{array}{cccccc} 2\lambda^+ \end{array} 
\\
&&&\\
\begin{array}{c}\chi^+  \end{array}
&\begin{array}{cccccc}  \phi\end{array} 
&\begin{array}{cccccc} 2\lambda^+\end{array} 
&\begin{array}{cccccc}  A^+ \end{array}
\\
\end{array}$
\caption{$D=4$, $[(\mathcal{N}=1)_{chiral}^{L}]\times[(\mathcal{N}=1)_{chiral}^{R}] = [(\mathcal{N}=4)_{SYM}]$.}
\label{squaring table for chiral multiplet}
\end{center}
\end{table} 
\FloatBarrier
The fields in the table can be organized traditionally into an $\mathcal{N}=2$ vector ($A^+,2\lambda^+,2\phi,2\lambda^-,A^-$) and an $\mathcal{N}=2$  hypermultiplet ($2\lambda^+,4\phi,2\lambda^-$). However, note that we have in fact obtained the field content of $\mathcal{N}=4$ SYM.

In the following subsection, we will show how to build the (on-shell) SUSY transformations of the $\mathcal{N}=4$ SYM multiplet from those of the $\mathcal{N}=1$ chiral multiplet; we will find that we need to introduce two extra generators for SUSY transformations (which are not symmetries) to achieve this\footnote{We will use a notation similar to \cite{Bianchi:2008pu} - there the authors  construct the SUSY transformations of $\mathcal{N}=8$ supergravity fields in terms of those of $\mathcal{N}=4$ SYM and give a dictionary between the lowering operators of the two theories.}. 


\subsection{Enhanced chiral multiplet and the extra SUSY generators}\label{Enhanced chiral multiplet and the extra SUSY generators-subsection}
We begin with the familiar position-space transformations of the $\mathcal{N}=1$ chiral multiplet (comprising of a left-handed Weyl fermion and a complex scalar):
\be
\label{chiral position space SUSY}
\delta_\epsilon\phi=\epsilon^a \chi_a,\qquad
\delta_\epsilon\chi_a=-i\sigma_{a\dot{b}}^\mu\epsilon^{\dagger\dot{b}}\partial_\mu \phi,
\ee
and similarly for $\bar{\phi}$ and $\chi_{\dot{a}}^\dagger$. We can write these in terms of 4-component Majorana spinors as:
\be
\label{susy chiral Majorana}
\delta_\epsilon\phi=\epsilon P_L\chi,\qquad
\delta(P_L\chi)=-iP_L(\gamma^\mu\partial_\mu\phi)\epsilon
\ee
and similarly for $\bar{\phi}$ and $P_R\chi$. It is useful to expand these fields in terms of creation and annihilation operators (since, as we see later, these will be the ones used in the dictionary):
\be
\begin{aligned}
\label{Free field expansions}
\phi(x)&=\int \widetilde{dp}[\phi_-(p)e^{ipx}+\phi_+^\dagger(p)e^{-ipx}], \\
P_L\chi(x)&=\sum_{s=\pm}\int\widetilde{dp}[\chi_s(p)(P_L u_s(p))_a e^{ipx}+\chi_s^\dagger(p)(P_Lv_s(p))_ae^{-ipx}],
\end{aligned}
\ee
where $\widetilde{dp}=\frac{d^3p}{(2\pi)^3 \sqrt{2E_p}}$ and $\phi_\pm$,$\chi_\pm$ satisfy the usual algebra of bosonic/fermionic creation and annihilation operators. Note that the $\pm$ labels on the bosonic operators $\phi$ tell us which of the fermionic operators they are related to via SUSY. 

\begin{figure}
\centering
\includegraphics[scale=0.20]{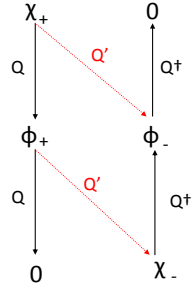}
\caption{Susy generators of the chiral multiplet}\label{fig:writing N=1 as N=2 diagram} 
\end{figure}

Given that we will be using the raising/lowering operators in our dictionary, we must describe how they transform under supersymmetry. The rules can be read off by combining \eqref{susy chiral Majorana} and \eqref{Free field expansions} and we get\footnote{For convenience, we will be using the notation of the spinor-helicity formalism; $|p]$ and $|p\rangle$ can be thought of as 2-component complex vectors satisfying the massless Weyl equations
\be
\begin{aligned}
p_{a\dot{b}}|p\rangle^{\dot{b}}&=0,\quad p_{a\dot{b}}=p_\mu(\sigma^\mu)_{a\dot{b}}=\bigl(\begin{smallmatrix}
p^0+p^3 &p^1-ip^2 \\ 
p^1+ip^2 &p^0-p^3 
\end{smallmatrix}\bigr)\\
p^{\dot{a}b}|p]_b&=0,\quad p^{\dot{a}b}=p_\mu(\bar{\sigma}^\mu)^{\dot{a}b}=\bigl(\begin{smallmatrix}
p^0-p^3 &-p^1+ip^2 \\ 
-p^1-ip^2 &p^0+p^3 
\end{smallmatrix}\bigr)
\end{aligned}
\ee
Note that sometimes the opposite convention for square and angle brackets is found in the literature; then one must swap $|p\rangle$ and $|p]$; here we are using the notation of \cite{Elvang:2013cua}. The relation between the familiar $u_s(p),v_s(p)$ and the complex vectors $|p]$ and $|p\rangle$ is:
\be
u_+(p)=v_-(p)=\binom{0}{|p\rangle^{\dot{a}}},\quad u_-(p)=v_+(p)=\binom{|p]_a}{0}
\ee

}:
\be
\label{var_states_original}
\begin{aligned}
\delta_\epsilon \chi_+(p)&=[\epsilon p] \phi_+(p)\\
\delta_\epsilon \phi_+(p)&=\langle \epsilon p\rangle \chi_+(p)\\
\delta_\epsilon \phi_-(p)&=[\epsilon p] \chi_-(p)\\
\delta_\epsilon \chi_-(p)&=\langle \epsilon p\rangle \phi_-(p)
\end{aligned}
\ee
It will be useful to write down the general form of the SUSY generators $Q_M=\binom{|Q]_a}{|Q\rangle^{\dot{a}}}$ as functions of $\phi$ and $\chi$. One can show that they are
\be
\begin{aligned}
[Q|^a&=\int\widetilde{dp}[p|^a(\phi_+(p)\chi_+^\dagger(p)-\chi_-(p)\phi_-^\dagger(p))\\
|Q^\dagger\rangle^{\dot{a}}&=\int\widetilde{dp}|p\rangle^{\dot{a}}(\phi_-(p)\chi_-^\dagger(p)-\chi_+(p)\phi_+^\dagger(p))
\end{aligned}
\ee
Then their action on the lowering operators is
\be
\begin{aligned}
\label{action of Q on lowering}
[Q,\chi_+(p)]&=[p|\phi_+(p)  \quad& [Q^\dagger,\chi_+(p)]&=0\qquad \\
[Q,\phi_+(p)]&=0  \quad &[Q^\dagger,\phi_+(p)]&=|p\rangle \chi_+(p)\qquad  \\
[Q,\phi_-(p)]&=[p|\chi_-(p)   \quad &[Q^\dagger,\phi_-(p)]&=0\qquad \\
[Q,\chi_-(p)]&=0   \quad &[Q^\dagger,\chi_-(p)]&=|p\rangle \phi_-(p)\qquad
\end{aligned}
\ee
where the bracket is to be understood as an anti-commutator when both arguments are Grassmanian.
In \cite{Bianchi:2008pu}, the eight SUSY generators of maximal supergravity came from two copies of the four maximal SYM generators. Although squaring the field content of the $\mathcal{N}=1$ chiral multiplet does give us the $\mathcal{N}=4$ SYM multiplet, it seems like we don't have enough SUSY generators to build the supersymmetry of the gauge multiplet. This can be resolved by noticing that our generator $Q$ only relates ($\phi_+$ and $\chi_+$) and ($\phi_-$ and $\chi_-$) separately. One can define a new generator $Q'$ which mixes ($\phi_+$ and $\chi_-$) and ($\phi_-$ and $\chi_+$) separately, as shown in \autoref{fig:writing N=1 as N=2 diagram}. 
Of course, for the purpose of describing the chiral multiplet, this extra generator is redundant, but we will see that it becomes crucial for obtaining the SUSY transformations of $\mathcal{N}=4$ SYM. One can apply an $SO(2)$ rotation on the bosonic states (keeping the fermionic states unchanged) to obtain a new set of SUSY generators:    
\be
\begin{aligned}
[Q'|^a&=\int\widetilde{dp}[p|^a (\phi_-(p) \chi_+^\dagger(p)+\chi_-(p)\phi_+^\dagger(p))\\
|Q'^\dagger\rangle^{\dot{a}}&=\int\widetilde{dp}|p\rangle^{\dot{a}} (-\phi_+(p)\chi_-^\dagger(p)-\chi_+(p)\phi_-^\dagger(p))
\end{aligned}
\ee
Note that this $SO(2)=U(1)$ is different from the $U(1)$ R-symmetry group, under which both the bosonic and fermionic states will transform\footnote{This can be seen for example from the interacting Lagrangian 
\be 
\mathcal{L}_I=\frac{1}{2}\phi\psi\psi+c.c.-\frac{1}{4}|g|^4|\phi|^4
\ee 
where it is obvious that $\phi$ must trasform with half the $U(1)$ charge of $\phi$.}. This is to be expected, since an R-symmetry rotation cannot give us new generators. 
Then the action of the new SUSY generators on the annihilation operators is (see \autoref{fig:writing N=1 as N=2 diagram}):
\be
\begin{aligned}
\label{action of Q'}
[Q',\chi_+(p)]&=[p|\phi_-(p) \quad &[Q'^\dagger,\chi_+(p)]&=0 \qquad\\ 
[Q',\phi_+(p)]&=-[p|\chi_-(p) \quad &[Q'^\dagger,\phi_+(p)]&=0 \qquad\\ 
[Q',\phi_-(p)]&=0 \quad &[Q'^\dagger,\phi_-(p)]&=|p\rangle \chi_+(p)  \qquad\\
[Q',\chi_-(p)]&=0 \quad &[Q'^\dagger,\chi_-(p)]&= -|p\rangle \phi_+(p) \qquad
\end{aligned}
\ee
We can thus rewrite the fields and SUSY generators in our $\mathcal{N}=1$ theory as
\be
\phi^i=\binom{\phi_+}{\phi_-},\quad
Q^i=\binom{Q}{Q'}
\ee
and it is straightforward to see that the fields will transform under the action of the SUSY generators as
\be
\label{action_of_enhanced_gen_on_states}
\begin{aligned}
[Q^i,\chi_+(p)]&=[p|\phi^i(p) \quad &[Q_i^\dagger,\chi_+]&=0 \qquad \\
[Q^i,\phi^j(p)]&=[p|\varepsilon^{ij}\chi_-(p) \quad &[Q_i^\dagger,\phi^j]&=|p\rangle \delta_i^j \chi_+(p) \qquad \\
[Q^i,\chi_-(p)]&=0 \quad &[Q_i^\dagger,\chi_-]&=|p\rangle\varepsilon_{ij}\phi^j(p) 
\end{aligned}
\ee
Some comments are in order. We note that \eqref{action_of_enhanced_gen_on_states} looks like the action of SUSY on an $\N=2$ multiplet. However, the states are still those of the $\N=1$ chiral multiplet. The point is that the action of $Q'$, despite giving legitimate\textit{ tranformations} on the momentum states, does not constitute a symmetry of the $\N=1$ Lagrangian. However, the remarkable thing is that it becomes a symmetry upon squaring (see \autoref{Dictionary chiral to super-Yang-Mills}). This sort of enhancement is not unusual when squaring multiplets, and has already been observed in \cite{Anastasiou:2015vba}. In that case SYM multiplets were tensored to obtain various supergravity theories. The R-symmetry algebra of the supegravity theory includes, but is not limited to, the two R-symmetry algebras of the gauge theories. The missing generators are understood as transformations of the SYM states, which become symmetries of the theory obtained from squaring. \\
Of course, one could have multiplied two $\N=2$ hypermultiplets; this would have given us four copies of the $\N=4$ SYM, transforming into each other under some global symmetry. Since our final goal was to show how $\N=8$ supergravity is constructed as a quadruple copy, we needed a single $\N=4$ multiplet, and this was achieved through the enhanced chiral. \\
Another upshot of this in-between construction is that, unlike the $\N=1$ chiral or the $\N=2$ hypermultiplet, it is actually CPT self-conjugate (this can be seen from \eqref{action_of_enhanced_gen_on_states} and \autoref{fig:writing N=1 as N=2 diagram}) and hence it naturally squares to a self- CPT conjugate multiplet, the $\N=4$ SYM.\\ 
Finally, we note that there are similarities between our construction and the half-hypermultiplet which appears elsewhere in the literature.\footnote{See, for example \cite{Chiodaroli:2015wal} or \cite{Bhardwaj:2013qia}}. The half-hypermultiplet has the same field content as the chiral; when its gauge group is in a pseudo-real representation over the direct sum of a complex representation $R$ and its conjugate, then it reduces to the usual hypermultiplet when considered over $R$. We have not made use of this trick here, as we would like to stay as general as possible with regards to the gauge group (see \autoref{Non-abelian gauge symmetry and R-symmetry}). 
 
\subsection{$\mathcal{N}=4$ SYM}\label{N=4 operators section}
On-shell, the maximal SYM multiplet (with R-symmetry $SU(4)$) consists of 1 gluon with helicity $h=+1$ ($B_\pm$), 4 gluinos with $h=+\frac{1}{2}$ ($\lambda_\pm^a$), 6 scalars with $h=0$ ($\phi^{ab}$), 4 gluinos with $h=-\frac{1}{2}$ ($\lambda_-^{abc}$) and 1 gluon with $h=-1$ ($A_-^{abcd}$), with $a,b,c,d=1,...4$ being the $SU(4)$ indices.  It is convenient to re-write the states as:
\be
A_+,\ \lambda_+^a,\ \phi^{ab}=\frac{1}{2}\varepsilon^{abcd}\phi_{cd},\ \lambda_a^-=\frac{1}{3!}\varepsilon_{abcd}\lambda_-^{bcd},\ A_-=-\frac{1}{4!}\varepsilon_{abcd}A^{abcd}
\ee
Then, under the action of the SUSY generators, the annihilation operators will transform as
\be
\begin{aligned}
\label{N=4 susy transformations operators}
[Q^a,A_+(p)]&=[p|\lambda_+^a(p) \quad &[Q_a^\dagger,A_+(p)]&=0 \qquad \\
[Q^a,\lambda_+^b(p)]&=[p|\phi^{ab}(p) \quad &[Q_a^\dagger,\lambda_+^b(p)]&=|p\rangle\delta_a^b A_+(p)   \qquad \\
[Q^a,\phi^{bc}(p)]&=[p|\varepsilon_-^{abcd}\lambda_d^-(p) \quad &[Q_a^\dagger,\phi^{bc}(p)]&=|p\rangle2!\delta_a^{[b}\lambda_+^{c]}(p) \qquad \\
[Q^a,\lambda_{b-}(p)]&=-[p|\delta_b^aA_-(p) \quad &[Q_a^\dagger,\lambda_{b-}(p)]&=|p\rangle\phi_{ab}(p) \qquad \\
[Q^a,A_-(p)]&=0 \quad &[Q_a^\dagger,A_-(p)]&=-|p\rangle\lambda^-_{a}(p) \qquad 
\end{aligned}
\ee

\subsection{Dictionary}\label{Dictionary chiral to super-Yang-Mills}
We now have all the ingredients necessary to build a dictionary between $\mathcal{N}=4$ SYM and two copies of the $\mathcal{N}=1$ (written as $\mathcal{N}=2$) chiral multiplet. We will split the $SU(4)$ indices $a,b,...=1,...4$ into left $SU(2)$ indices ($i,j,...=1,2$) and right $SU(2)$ indices $r,s...=1,2$. The SYM operators can then be built as  tensor products of the form $O_L\otimes \tilde{O}_R$, with $Q^i$ and $Q^r$ acting only on the LHS and RHS respectively. We can then write a dictionary for all the operators described in \autoref{N=4 operators section}:
\be
\begin{aligned}
\label{dictionary chiral SYM}
A_+(p)&=\chi_+(p)\otimes \tilde{\chi}_+(p)\\
\lambda_+^a(p)&=
\left\{\begin{matrix}
\lambda_+^i(p)=\phi^i(p)\otimes\tilde{\chi}_+(p)\\ 
\lambda_+^r(p)=\chi_+(p)\otimes\tilde{\phi}^r(p)
\end{matrix}\right. \\
\phi^{ab}(p)&=
\left\{\begin{matrix}
\phi^{ij}(p)=\varepsilon^{ij}\chi_-(p)\otimes\tilde{\chi}_+(p)\\ 
\phi^{ir}(p)=\phi^i(p)\otimes\tilde{\phi}^r(p)\ \ \ \ \ \\ 
\phi^{rs}(p)=\varepsilon^{rs}\chi_+(p)\otimes\tilde{\chi}_-(p)
\end{matrix}\right.\\
\lambda_a^-(p)&=
\left\{\begin{matrix}
\lambda_i^-(p)=\phi_i(p)\otimes\tilde{\chi}_-(p)\\ 
\lambda_r^-(p)=\chi_-(p)\otimes\tilde{\phi}_r(p)
\end{matrix}\right. \\
A_-(p)&=\chi_-(p) \tilde{\chi}_-(p)\\
\end{aligned}
\ee
It has been checked that for any SYM operator, its SUSY transformation can be obtained from those of the  LHS and RHS chiral multiplets.

As an interesting example, let's look at the action of $Q_a^\dagger$ on the scalars $\phi^{ab}(p)$. We know from \eqref{N=4 susy transformations operators} that
\be
[Q_a^\dagger,\phi^{bc}(p)]=|p\rangle (\delta_a^b\lambda_+^c(p)-\delta_a^c \lambda_+^b(p))
\ee
We now decompose this in terms of the LHS and RHS indices to get the following set of transformation rules:
\begin{itemize}
\item all indices LHS, we expect

\be
[Q_i^\dagger,\phi^{jk}(p)]=|p\rangle (\delta_i^j\lambda_+^k(p)-\delta_i^k \lambda_+^j(p))
\ee
To check this, we substitute the dictionary \eqref{dictionary chiral SYM} and get
\be
\begin{aligned}
[Q_i^\dagger,\varepsilon^{jk}\chi_-(p)\otimes\tilde{\chi}_+(p)]&=\varepsilon^{jk}[Q_i^\dagger,\chi_-(p)]\otimes\tilde{\chi}_+(p)\\
&=|p\rangle\varepsilon^{jk}\varepsilon_{il} \phi^l(p)\otimes\tilde{\chi}_+(p)\\
&=|p\rangle (\delta_i^j\lambda_+^k(p)-\delta_i^k \lambda_+^j(p))
\end{aligned}
\ee  
as expected.

\item two LHS indices, one RHS index, we expect

\be
[Q_i^\dagger,\phi^{jr}]=|p\rangle (\delta_i^j \lambda_+^r)
\ee
We check this using the dictionary:
\be
\begin{aligned}
[Q_i^\dagger,\phi^j(p)\otimes\tilde{\phi}^r(p)]&=[Q_i^\dagger,\phi^j]\otimes\tilde{\phi}^r \\
&=|p\rangle\delta_i^j \chi_+(p)\otimes\tilde{\phi}^r(p)\\
&=|p\rangle\delta_i^j \lambda_+^r(p)
\end{aligned}
\ee
as expected. We also have
\be
[Q_r^\dagger,\phi^{ij}(p)]=0
\ee
and we can again check 
\be
[Q_r^\dagger,\varepsilon^{ij}\chi_-(p)\otimes\tilde{\chi}_+(p)]=\varepsilon^{ij}\chi_-(p)\otimes[Q_r^\dagger,\tilde{\chi}_+]=0
\ee
\end{itemize}
All the other cases will proceed similarly to the ones above.

\subsection{Non-abelian gauge symmetry and R-symmetry}\label{Non-abelian gauge symmetry and R-symmetry}

We can allow the states of our $\mathcal{N}=1$ multiplets to transform in the adjoint of some non-abelian groups $G_{L/R}$:
\be
\delta \Phi_{L/R}^\alpha=f^\alpha_{\ \beta\gamma}\Phi^\beta \theta^\gamma
\ee
with $\alpha=1,2,\ldots,\text{dim}(G_{L/R})$ and $\theta^\alpha$ a global parameter. We will tensor these to obtain linearized SYM, where the abelian local $U(1)$ transformations are decoupled from the non-abelian global ones. Under the latter, the SYM fields will also transform in the adjoint representation and this suggests the natural dictionary:
\be
\Phi_{SYM}^\alpha=f^\alpha_{\ \beta\gamma}\Phi_L^\beta\Phi_R^\gamma
\ee
which requires $G_L=G_R=G_{SYM}$. Given the absence of local gauge parameters in the chiral multiplet transformations, we see that it will be the field strength, rather than the potential, that is obtained through squaring.\\

The next question is how the $SU(4)$ R-symmetry is built from transformations of the $\mathcal{N}=1$ fields. First we notice that after introducing the extra SUSY generators, the automorphism of the SUSY algebra, whose generators act on the Q's via
\be
[T_A,Q_{a}]=(U_A)_a{}^b Q_{b}
,\qquad a, b =1,\ldots,\N \ee
 is enhanced to $SU(2)$ for our faux $\mathcal{N}=2$ multiplet\footnote{Note that it is $SU(2)$ rather than $U(2)$ because the introduction of the extra SUSY generator makes the enhanced $\mathcal{N}=1$ multiplet CPT conjugate. Interestingly, it seems that one can only obtain a CPT self-conjugate multiplet by tensoring two multiplets which are CPT conjugate themselves. This is also observed at the next level of squaring, when we get $\mathcal{N}=8$ Supergravity from squaring $\mathcal{N}=4$ SYM.}. 

We note that the situation is similar to what we had before for $Q'$ (see \autoref{Enhanced chiral multiplet and the extra SUSY generators-subsection}): $SU(2)$ is not a symmetry of our $N=1$ theory, when thought of as a stand-alone chiral multiplet; it becomes a symmetry only when acting on tensor products (i.e. after squaring). The generators can be built as $(T^a_b)^c_{\ d}=\delta^a_d\delta^c_b -\frac{1}{2}\delta^a_b\delta^c_d$ and $a,b,c,d=1,2$\footnote{We use the same conventions as in \cite{Bianchi:2008pu}.}. For example, on a subset of the spinors, these will act as:
\be
\label{fundamental su2s}
[T^i_{\ j},\phi^k\otimes\tilde{\chi}]=[T^i_{\ j},\lambda^k]=\delta^k_j\lambda^i
\ee

 Then $SU(4)$ will be built from $SU(2)_L\times SU(2)_R \times U(1)$: we will have $3_L+3_R$ $SU(2)$ generators acting as in \eqref{fundamental su2s}, and a $U(1)$ generator $T$ which acts on the SYM states via:
\be
\begin{aligned}
[T,A_+]&=0\\
[T,\lambda_+^i]&=\lambda_+^i,\ [T,\lambda_+^r]=-\lambda_+^r\\
[T,\phi^{ij}]&=2\phi^{ij},\ [T,\phi^{rs}]=-2\phi^{rs},\ [T,\phi^{ir}]=0
\end{aligned}
\ee
In addition, we have the generators which mix the LHS and RHS states. They are given by $(T^a_r)^s_{\ b}=\delta^a_b\delta_r^s$ and $(T^r_a)^b_{\ s}=\delta^b_a\delta_s^r$ and their action on the SYM states is
\be
[T^r_{\ b},\phi_{is}]=\delta_s^r\phi_{ib}
\ee
Note that each of the mixed generators can be interpreted as a tensor product of the SUSY operators, where we have formally supressed the spacetime indices, for example $T^s_{\ a}=q^s\otimes q_a$, where $[q_a,\phi_b]=\varepsilon_{ab}\lambda_-$\footnote{Interestingly, the same interpretation can be given to the terms that mix the LHS and RHS in the global symmetry of supergravities obtained from SYM squaring \cite{Anastasiou:2015vba}.}.
In conclusion, the action of the R-symmetry generators on the SYM states was built  from the action on the states of the  chiral multiplet in a similar way to the action on supergravity states from the action on SYM states\cite{Bianchi:2008pu}.

Interestingly, the R symmetry of this theory is then built via the same formula that we gave in \cite{Anastasiou:2015vba} for $\mathfrak{h}$, where $\mathfrak{h}$ is the maximally compact subalgebra of the U-duality of supergravity theories obtained from squaring\footnote{In \cite{Anastasiou:2015vba}, we actually find the general formula for any $3\leq D \leq 10$:
\be
\label{R formula}
\mathfrak{sa}(\mathcal{N}_L+\mathcal{N}_R, \mathds{D})=\big[\mathfrak{sa}(\mathcal{N}_L, \mathds{D})\oplus\mathfrak{sa}(\mathcal{N}_R, \mathds{D})\oplus\delta_{D,4}\mathfrak{u}(1)+\mathds{D}[\mathcal{N}_L, \mathcal{N}_R]\big]
\ee
where $\mathds{D}$ is the division algebra associated with the spinor representation in dimension $D$ and $\mathfrak{sa}(n,\alg)\cong\mathfrak{Isom}(\alg \mathds{P}^{n-1})$.}:
\be
\begin{aligned}
\mathfrak{su}(4)&=\mathfrak{su}(2)\oplus\mathfrak{su}(2)\oplus\mathfrak{u}(1)+\C[2,2]\\
&=\mathfrak{r}_L\oplus\mathfrak{r}_R\oplus\mathfrak{u}(1)+q_L\otimes q_R
\end{aligned}
\ee
and $q_{L/R}$ are obtained from the supercharges as explained above.

\subsection{Gravity-chiral dictionary}
On-shell, the maximal Supergravity multiplet consists of a graviton, 8 gravitini, 28 vectors, 56 fermions and 70 scalars, whose helicity states are represented as:
\be
g_+,\ \psi_+^A,\ A_+^{AB},\ \chi_+^{ABC},\ \phi^{ABCD},\ \chi^-_{ABC},\ A^-_{AB},\ \psi^-_A,\ g^-
\ee
Then the gravity-chiral dictionary is \footnote{Note that the RHS SYM fields are dashed. We use the following convention for the indices:
\begin{itemize}
\item supergravity: A, B, C...
\item SYM: a,b,...(LHS) and $\bar{a}$,$\bar{b}$,... (RHS)  
\item chiral: i($\bar{i}$),j,($\bar{j}$)...(LHS) and r($\bar{r}$),s($\bar{s}$),... (RHS)
\end{itemize}} (via $\mathcal{N}=4$ SYM):

\begin{align*}\label{grav chiral dictionary}
g&=A_+A_-'=\chi_+\tilde{\chi}_+\chi_-' \tilde{\chi}_-'  \numberthis\\
\psi^A&=
\left\{\begin{matrix}
\psi_+^a=\lambda_+^aA_+'=
\begin{Bmatrix}
\lambda_+^i=\phi^i\tilde{\chi}_+\\ 
\lambda_+^r=\chi_+\tilde{\phi}^r
\end{Bmatrix}
\chi_+' \tilde{\chi}_+'\\ 
\psi_+^{\bar{a}}=A_+(\lambda')_+^{\bar{a}}=\chi_+\tilde{\chi}_+
\left\{\begin{matrix}
(\lambda')_+^{\bar{i}}=(\phi')^{\bar{i}}\tilde{\chi'}_+\\ 
(\lambda')_+^{\bar{r}}=\chi'_+(\tilde{\phi'})^{\bar{r}}
\end{matrix}\right.
\end{matrix}\right.\\
A_+^{AB}&=
\left\{\begin{matrix}
A^{ab}_+=\phi^{ab}A'_+=
\begin{Bmatrix}
\phi^{ij}=\varepsilon^{ij}\chi_-\tilde{\chi}_+\\ 
\phi^{ir}=\phi^i\tilde{\phi}^r\ \ \ \ \ \\ 
\phi^{rs}=\varepsilon^{rs}\chi_+\tilde{\chi}_-
\end{Bmatrix}\chi_+' \tilde{\chi}_+'\\
A^{a\bar{a}}_+=\lambda^a_+(\lambda')^{\bar{a}}_+ =
\begin{Bmatrix}
\lambda_+^i=\phi^i\tilde{\chi}_+\\ 
\lambda_+^r=\chi_+\tilde{\phi}^r
\end{Bmatrix}
\left\{\begin{matrix}
(\lambda')_+^{\bar{i}}=(\phi')^{\bar{i}}\tilde{\chi'}_+\\ 
(\lambda')_+^{\bar{r}}=\chi'_+(\tilde{\phi'})^{\bar{r}}\\
\end{matrix}\right.\\
A^{\bar{a}\bar{b}}_+=A_+(\phi')^{\bar{a}\bar{b}}=\chi_+\tilde{\chi}_+
\left\{\begin{matrix}
(\phi')^{\bar{i}\bar{j}}=\varepsilon^{\bar{i}\bar{j}}\chi'_-(\tilde{\chi})'_+\\ 
(\phi')^{\bar{i}\bar{r}}=(\phi')^{\bar{i}}(\tilde{\phi}')^{\bar{r}}\ \ \ \ \ \\ 
(\phi')^{\bar{r}\bar{s}}=\varepsilon^{\bar{r}\bar{s}}\chi'_+(\tilde{\chi})'_-
\end{matrix}\right.
\end{matrix}\right.\\
\chi_+^{ABC}&=
\left\{\begin{matrix}
\chi_+^{abc}=\varepsilon^{abcd}\lambda_d^-A'_+=\varepsilon^{abcd}
\begin{Bmatrix}
\lambda_i^-=\phi_i\tilde{\chi}_-\\ 
\lambda_r^-=\chi_-\tilde{\phi}_r
\end{Bmatrix}
\chi_+' \tilde{\chi}_+'\\ 
\chi_+^{ab\bar{a}}=\phi^{ab}(\lambda')_+^{\bar{a}}=
\begin{Bmatrix}
\phi^{ij}=\varepsilon^{ij}\chi_-\tilde{\chi}_+\\ 
\phi^{ir}=\phi^i\tilde{\phi}^r\ \ \ \ \ \\ 
\phi^{rs}=\varepsilon^{rs}\chi_+\tilde{\chi}_-
\end{Bmatrix}
\left\{\begin{matrix}
(\lambda')_+^{\bar{i}}=(\phi')^{\bar{i}}\tilde{\chi'}_+\\ 
(\lambda')_+^{\bar{r}}=\chi'_+(\tilde{\phi'})^{\bar{r}}\\
\end{matrix}\right.\\ 
\chi_+^{a\bar{a}\bar{b}}=\lambda_+^a(\phi')^{\bar{a}\bar{b}}=
\begin{Bmatrix}
\lambda_+^i=\phi^i\tilde{\chi}_+\\ 
\lambda_+^r=\chi_+\tilde{\phi}^r
\end{Bmatrix}
\left\{\begin{matrix}
(\phi')^{\bar{i}\bar{j}}=\varepsilon^{\bar{i}\bar{j}}\chi'_-(\tilde{\chi})'_+\\ 
(\phi')^{\bar{i}\bar{r}}=(\phi')^{\bar{i}}(\tilde{\phi}')^{\bar{r}}\ \ \ \ \ \\ 
(\phi')^{\bar{r}\bar{s}}=\varepsilon^{\bar{r}\bar{s}}\chi'_+(\tilde{\chi})'_-
\end{matrix}\right.\\ 
\chi_+^{\bar{a}\bar{b}\bar{c}}=\varepsilon^{\bar{a}\bar{b}\bar{c}\bar{d}}A_+(\lambda')_{\bar{d}}^-=
\varepsilon^{\bar{a}\bar{b}\bar{c}\bar{d}}\chi_+\tilde{\chi}_+
\left\{\begin{matrix}
(\lambda')_{\bar{i}}^-=\phi'_{\bar{i}}(\tilde{\chi})'_-\\ 
(\lambda')_{\bar{r}}^-=\chi'_-(\tilde{\phi})'_{\bar{r}}
\end{matrix}\right.
\end{matrix}\right. \\
\phi^{ABCD}&=
\left\{\begin{matrix}
\phi^{abcd}=\varepsilon^{abcd}A_-A'_+=
\chi_-\tilde{\chi}_-
\chi_+' \tilde{\chi}_+'\\ 
\phi^{abc\bar{a}}=\varepsilon^{abcd}\lambda_d^-(\lambda')_+^{\bar{a}}=
\varepsilon^{abcd}
\begin{Bmatrix}
\lambda_i^-=\phi_i\tilde{\chi}_-\\ 
\lambda_r^-=\chi_-\tilde{\phi}_r
\end{Bmatrix}
\left\{\begin{matrix}
(\lambda')_+^{\bar{i}}=(\phi')^{\bar{i}}\tilde{\chi'}_+\\ 
(\lambda')_+^{\bar{r}}=\chi'_+(\tilde{\phi'})^{\bar{r}}\\
\end{matrix}\right.\\ 
\phi^{ab\bar{a}\bar{b}}=\phi^{ab}(\phi')^{\bar{a}\bar{b}}=
\begin{Bmatrix}
\phi^{ij}=\varepsilon^{ij}\chi_-\tilde{\chi}_+\\ 
\phi^{ir}=\phi^i\tilde{\phi}^r\ \ \ \ \ \\ 
\phi^{rs}=\varepsilon^{rs}\chi_+\tilde{\chi}_-
\end{Bmatrix}
\left\{\begin{matrix}
(\phi')^{\bar{i}\bar{j}}=\varepsilon^{\bar{i}\bar{j}}\chi'_-(\tilde{\chi})'_+\\ 
(\phi')^{\bar{i}\bar{r}}=(\phi')^{\bar{i}}(\tilde{\phi}')^{\bar{r}}\ \ \ \ \ \\ 
(\phi')^{\bar{r}\bar{s}}=\varepsilon^{\bar{r}\bar{s}}\chi'_+(\tilde{\chi})'_-
\end{matrix}\right.\\ 
\phi^{a\bar{a}\bar{b}\bar{c}}=\varepsilon^{\bar{a}\bar{b}\bar{c}\bar{d}}\lambda_+^a(\lambda')_{\bar{d}}^-=
\varepsilon^{\bar{a}\bar{b}\bar{c}\bar{d}}
\begin{Bmatrix}
\lambda_+^i=\phi^i\tilde{\chi}_+\\ 
\lambda_+^r=\chi_+\tilde{\phi}^r
\end{Bmatrix}
\left\{\begin{matrix}
(\lambda')_{\bar{i}}^-=\phi'_{\bar{i}}(\tilde{\chi})'_-\\ 
(\lambda')_{\bar{r}}^-=\chi'_-(\tilde{\phi})'_{\bar{r}}
\end{matrix}\right.\\ 
\phi^{\bar{a}\bar{b}\bar{c}\bar{d}}=\varepsilon^{\bar{a}\bar{b}\bar{c}\bar{d}}A_+(A')_-=
\varepsilon^{\bar{a}\bar{b}\bar{c}\bar{d}}\chi_+\tilde{\chi}_+\chi_-' \tilde{\chi}_-'
\end{matrix}\right.
\end{align*}
and similarly for the negative helicity states.

\section{Squaring in $D\neq 4$} \label{Squaring in D neq 4}
In 6 dimensions, we would like to build the maximal $\mathcal{N}=(2,2)$ supergravity out of 4 copies of the chiral multiplet. The easiest route is to first write the $\mathcal{N}=(2,0)$ tensor multiplet as two copies of the $\mathcal{N}=(1,0)$ multiplet (and similarly for $\mathcal{N}=(0,2)$). Then it is straightforward to combine the two tensor multiplets and obtain supergravity.

The $\mathcal{N}=(1,0)$ on-shell chiral superfield is
\be
\chi(\eta^{i+})=\chi^++\phi_i\eta^{i+}+\eta^{i+}\eta^{j+}\Omega_{ij}\chi^-
\ee 
Note that we take $\eta^{ia}\to\eta^{i+}$ in order to construct the superspace, breaking the $SU(2)$ little group symmetry (see \cite{Huang:2010rn}).
Our states are $\chi^+$, $\phi_i$ and $\chi^-$, where $\pm$ are the $SU(2)$ weights and $i=1,2$ is the $Sp(1)$ R-symmetry index. Then the action of the SUSY generator on the states is given by
\be
\begin{aligned}
\{ q^A_{i+},\chi^+\}&=\lambda_+^A\phi_i\\
\{ q^A_{i+}, \phi_j\}&=\lambda_+^A\Omega_{ij}\chi^-\\
\{ q^A_{i+},\chi^-\}&=0
\end{aligned}
\ee
where $A=1,...4$ are $SO(6)=SU(4)$ Lorentz indices in $D=6$ and $\lambda_+^A$ is the $+$ component of the $\lambda^{Aa}$ solution of the Weyl equation in $D=6$ (see \cite{Cheung:2009dc} for a description of the on-shell spinor-helicity formalism in six dimensions).

Now we want to build the $\mathcal{N}=(2,0)$ tensor multiplet out of two copies of the chiral multiplet above. The tensoring table in terms of the little group representations is given by:
\FloatBarrier
\begin{table}[h]
\begin{center}
 $\begin{array}{c|c|c}
&\begin{array}{c}  \chi \\(\rep{2},\rep{1})\end{array}& \begin{array}{c}2  \phi \\ 2(\rep{1},\rep{1}) \end{array}\\
\hline
&&\\
\begin{array}{c}\tilde{\chi}\\ (\rep{2},\rep{1})\end{array}
&\begin{array}{cccccc} B_{\mu\nu}+\phi  \\ (\rep{3},\rep{1})+ (\rep{1},\rep{1})\end{array} 
&\begin{array}{cccccc} 2\psi \\ 2(\rep{2},\rep{1})\end{array} 
\\
&&\\
\begin{array}{c}2 \phi \\ 2(\rep{1},\rep{1}) \end{array}
&\begin{array}{cccccc} 2\psi \\ 2(\rep{2},\rep{1}) \end{array} 
&\begin{array}{cccccc} 4\phi\\ 4(\rep{1},\rep{1}) \end{array} 

\end{array}$
\caption{$D=6$, $[(1,0)_{chiral}^{L}]\times[(1,0)_{chiral}^{R}] = [(2,0)_{\text{\emph{Tensor}}}]$.}\label{tensor from chiral table}
\end{center}
\end{table}
\FloatBarrier
\noindent
The on-shell superfield for the $\mathcal{N}=(2,0)$ theory is given by \cite{Huang:2010rn}
\be
\Phi(\eta^{I+})=B^++\psi^+_I\eta^{+I}+\frac{1}{2}\eta^{+I}\eta^{+J}[\phi_{IJ}+\Omega_{IJ}A^0]+\frac{1}{3!}\epsilon_{LIJK}\eta^{+I}\eta^{+J}\eta^{+K}\psi^{-L}+(\eta^+)^4 B^-
\ee
so, in descending order of $SU(2)$ weight, our states are
\be
B^+,\ \ \psi^+_I,\ \ [\phi_{IJ}+\Omega_{IJ}A^0]\equiv A_{IJ},\ \ \psi^{-I},\ \ B^-
\ee
where $I,J=1,...4$ are the $Sp(2)$ R-symmetry indices.
The action of the SUSY generators on these states is given by
\be
\begin{aligned}
\{q_I^A , B^+\}&=\lambda^A_+ \psi^+_I \\
\{q_I^A , \psi^+_J\}&=\lambda^A_+ A_{IJ} \\
\{q_I^A , A_{JK}\}&=\lambda^A_+ \varepsilon_{LIJK}\psi^{-L} \\
\{q_I^A , \psi^{-J}\}&=\lambda^A_+ \delta_I^J B^- \\
\{q_I^A , B^-\}&=0
\end{aligned}
\ee
Then one can write a dictionary for the tensor multiplet as a double copy of the chiral one:
\be
\begin{aligned}
B^+&=\chi^+\otimes \tilde{\chi}^+ \\
\psi^+_I&=\left\{\begin{matrix}
\psi^+_i=\phi_i\otimes \tilde{\chi}^+\\ 
\psi^+_r=\chi\otimes \tilde{\phi}_r  
\end{matrix}\right. \\
A_{IJ}&=\left\{\begin{matrix}
A_{ij}=\Omega_{ij}\chi^{-}\otimes\tilde{\chi}^+\\ 
A_{ir}=\phi_i\otimes\tilde{\phi}_r\ \ \ \ \ \ \  \\ 
A_{rs}=\Omega_{rs}\chi^+\otimes\tilde{\chi}^-
\end{matrix}\right. \\
\psi^{-I}&=\left\{\begin{matrix}
\psi^{-i}=\Omega^{ij}\phi_j\otimes\tilde{\chi}^-\\ 
\psi^{-r}=\Omega^{rs}\chi^-\otimes\tilde{\phi}_s
\end{matrix}\right. \\
B^-&=\chi^-\otimes \tilde{\chi}^- 
\end{aligned}
\ee
It has been checked that the SUSY transformations of the tensor multiplet then follow from those of the chiral multiplet, similarly to the situation in $D=4$.
The R-symmetry is again obtained via the formula \eqref{R formula} 
\be
\begin{aligned}
\mathfrak{sp}(2)&=\mathfrak{sp}(1)\oplus\mathfrak{sp}(1)+\Q[1,1]\\
&=\mathfrak{r}_L\oplus\mathfrak{r}_R+q_L\otimes q_R
\end{aligned}
\ee
noting that the representation in D=6 is quaternionic. Here $\mathfrak{sp}(1)=\mathfrak{su}(2)$ is the R-symmetry algebra of the left and right chiral multiplet, and again the total R-symmetry of the resulting multiplet is enhanced via a tensor product of fermionic transformations (which are not symmetries) of the $\mathcal{N}=1$ states.

We can now perform the more straightforward squaring of the two tensor multiplets of opposite chiralities to get the maximal supergravity, as shown in the table below:       

\FloatBarrier
\begin{table}[h]
\begin{center}
 $\begin{array}{c|c|c|c}
&\begin{array}{c}  {B}^+_{\mu\nu} \\(\rep{3},\rep{1})\end{array}& \begin{array}{c}4  \lambda^{+} \\ 4(\rep{2},\rep{1}) \end{array}&\begin{array}{c} 5 \phi \\ (\rep{1},\rep{1})\end{array}\\

\hline
&&&\\
\begin{array}{c}B^-_{\mu\nu}\\ (\rep{1},\rep{3})\end{array}
&\begin{array}{cccccc} g_{\mu\nu}  \\ (\rep{3},\rep{3})\end{array} 
&\begin{array}{cccccc} 4\Psi_\mu^+ \\ 4(\rep{2},\rep{3})\end{array} 
&\begin{array}{cccccc} 5 B^-_{\mu\nu}\\ 5(\rep{1},\rep{3}) \end{array} 
\\

&&&\\
\begin{array}{c}4 {\lambda}^{-} \\ 4(\rep{1},\rep{2}) \end{array}
&\begin{array}{cccccc} 4\Psi^{-} \\ 4(\rep{3},\rep{2}) \end{array} 
&\begin{array}{cccccc} 16A_\mu\\ 16(\rep{2},\rep{2}) \end{array} 
&\begin{array}{cccccc} 20 \chi^{-} \\ 20(\rep{1},\rep{2})\end{array} 

\\

&&&\\

\begin{array}{c}5 \phi \\ 5(\rep{1},\rep{1})\end{array}
&\begin{array}{cccccc} 5 B^+_{\mu\nu}\\ 5(\rep{3},\rep{1}) \end{array} 
&\begin{array}{cccccc} 20 \chi^{+} \\ 20(\rep{2},\rep{1})\end{array} 
&\begin{array}{cccccc} 25\varphi \\ (\rep{1},\rep{1})\end{array} 

\end{array}$
\caption{$D=6$, $[(2,0)_{tensor}^{L}]\times[(0,2)_{tensor}^{R}] = [(2,2)_{\text{\emph{SuGra}}}]$.}\label{trivially maximal SuGra}
\end{center}
\end{table}
\FloatBarrier
\noindent
Note that one could have alternatively built the $\mathcal{N}=(1,1)$ SYM multiplet (consisting of a gauge field transforming in the $\rep{(2,2)}$ of the little group, two spinors $\lambda^+$ and two $\lambda^-$, transforming in the $\rep{(2,1)}$ and $\rep{(1,2)}$ representations, respectively, and 4 scalars) by tensoring the $\mathcal{N}=(1,0)$ and $\mathcal{N}=(0,1)$ chiral multiplets. In five dimensions, where the little group is $SO(3)$, the tensoring will proceed  as detailed in \autoref{tensor from chiral table}, by simply supressing the $SO(3)$ subgroup under which the fields transform trivially.

\section{Conclusions} 
Having constructed the states and symmetries of gauge theories by squaring, one can hope to apply them to simplifying scattering amplitudes. We illustrate the appearance of the double copy structure for 4-point gluon scattering through an example. In the $\N=1$ chiral theory, the $s$-channel 4-point scattering process for two negative-helicity and two positive helicity fermions in Yukawa theory is given by (up to the coupling constant)
\be
A_4(1^-,2^-,3^+,4^+)=A^s_4(\bar{f}_1^-f_2^-\bar{f}_3^+f_4^+)\propto\frac{\langle 12\rangle}{\langle 34\rangle}
\ee
Since the 3-point fermion scattering process is not allowed, this is the smallest possible amplitude. We note that, in $\N=4$ SYM the scattering amplitude of the four gluons obtained by taking two copies of the fermions above is
\be
\label{amplitude fermi squared}
A_4(g_1^-g_2^-g_3^+g_4^+)\propto \frac{\langle 12\rangle^4}{\langle 12\rangle \langle 23\rangle\langle 34\rangle\langle 41\rangle}=- \frac{s_{12}}{s_{23}}\left ( \frac{\langle 12\rangle}{\langle 34\rangle}\right )^2=- \frac{s_{12}}{s_{23}}[A^s_4(\bar{f}_1^-f_2^-\bar{f}_3^+f_4^+)]^2
\ee
 after some rearrangements and after imposing momentum conservation on the 4 legs (we use $s_{ij}=(p_i+p_j)^2=\langle ij\rangle[ij]$).

 One could potentially find similar relations for a higher number $n$ of external legs, provided n is even, i.e. when the fermion amplitudes are non-vanishing. 

We can also now write the gravitational 4-point amplitude as four copies of the chiral one via
\be
A_4(h^-_1h^-_2h^+_3h^+_4)=-s_{12} A_4 (g_1^- g_2^- g_3^+g_4^+) A_4 (g_1^-g_2^-g_4^+g_3^+)\propto -\frac{(s_{12})^3}{s_{23}s_{24}} [A^s_4(\bar{f}_1^-f_2^-\bar{f}_3^+f_4^+)]^4
\ee 
Again, we expect similar (though more complicated) quadruple copy relations for a higher number of external legs. \footnote{It is not clear at this stage how the 3-point gluon amplitude in $\N=4$ SYM could be built as a fermionic double copy. One might ask instead whether we can get $A_3 (g_1^- g_2^- \phi_3) $ in $\N=4$ SYM as a double copy of the amplitude $A_3 (f_1^- f_2^- \phi_3 )$ from the chiral theory. We note that the double copy includes a ratio of Mandelstam variables (see \eqref{amplitude fermi squared}) this is expected to hold at 3 points as well, and, since $s_{ij}$ vanish there, this will present an obstruction to making predictions at three points.} 

It would also be interesting to see if there are any connections with (some modified version of) the BCJ relations. Additionally, one can make use of the SUSY Ward identities, in conjunction with the gravity-chiral dictionary in \eqref{grav chiral dictionary}, to find a variety of other quadruple copy relations. Note that to achieve this, we will need to make use of the additional $Q'$ generators introduced in \autoref{Enhanced chiral multiplet and the extra SUSY generators-subsection}. In fact, following the appearance of this note, new work has appeared \cite{Schreiber:2016sss} which employs chiral squaring in the context of the KLT relations. 

In this note we have restricted ourselves to the on-shell symmetries of the free theories. A small modification of the transformations \eqref{chiral position space SUSY} can render them appropriate for an interacting theory. However, this would also make them non-linear and to restore linearity one must go off-shell. Then one could investigate whether something can be inferred about the off-shell closure of the SUSY algebra of $\mathcal{N}=4$ SYM from its double copy structure. Note that in \cite{Anastasiou:2014qba} we tensored off-shell (super)fields to obtain an off-shell supergravity superfield.  It would be interesting to explore whether there are any connections with recent attempts at $\mathcal{N}=4$ off-shell SUSY via lower-dimensional ``holograms" \cite{Calkins:2014exa}.\\

\section*{Acknowledgments}
I am grateful to my PhD supervisor Mike Duff and to Leron Borsten for useful suggestions. I have also enjoyed fruitful discussions with A. Anastasiou and M. Hughes. I would also like to thank M. Bianchi for useful discussions on scattering amplitudes. I was supported by an EPSRC grant and CAMGSD, project UID/EEI/0009/2015. 

\newpage

\bibliographystyle{utphys}
\bibliography{Ref_Library}

\providecommand{\href}[2]{#2}\begingroup\raggedright\begin{thebibliography}{10}

\bibitem{kawai:1985xq}
H.~Kawai, D.~Lewellen, and S.~Tye, ``{A Relation Between Tree Amplitudes of
  Closed and Open Strings},''
\href{http://dx.doi.org/10.1016/0550-3213(86)90362-7}{{\em Nucl.Phys.}
  {\bfseries B269} (1986) 1}.

\bibitem{Bern:2008qj}
Z.~Bern, J.~Carrasco, and H.~Johansson, ``{New Relations for Gauge-Theory
  Amplitudes},'' \href{http://dx.doi.org/10.1103/PhysRevD.78.085011}{{\em
  Phys.Rev.} {\bfseries D78} (2008) 085011},
\href{http://arxiv.org/abs/0805.3993}{{\ttfamily arXiv:0805.3993 [hep-ph]}}.

\bibitem{Bern:2010ue}
Z.~Bern, J.~J.~M. Carrasco, and H.~Johansson, ``{Perturbative Quantum Gravity
  as a Double Copy of Gauge Theory},''
  \href{http://dx.doi.org/10.1103/PhysRevLett.105.061602}{{\em Phys.Rev.Lett.}
  {\bfseries 105} (2010) 061602},
\href{http://arxiv.org/abs/1004.0476}{{\ttfamily arXiv:1004.0476 [hep-th]}}.

\bibitem{Bern:2010yg}
Z.~Bern, T.~Dennen, Y.-t. Huang, and M.~Kiermaier, ``{Gravity as the Square of
  Gauge Theory},'' \href{http://dx.doi.org/10.1103/PhysRevD.82.065003}{{\em
  Phys.Rev.} {\bfseries D82} (2010) 065003},
\href{http://arxiv.org/abs/1004.0693}{{\ttfamily arXiv:1004.0693 [hep-th]}}.

\bibitem{Bianchi:2008pu}
M.~Bianchi, H.~Elvang, and D.~Z. Freedman, ``{Generating Tree Amplitudes in N=4
  SYM and N = 8 SG},''
  \href{http://dx.doi.org/10.1088/1126-6708/2008/09/063}{{\em JHEP} {\bfseries
  0809} (2008) 063},
\href{http://arxiv.org/abs/0805.0757}{{\ttfamily arXiv:0805.0757 [hep-th]}}.

\bibitem{Huang:2012wr}
Y.-t. Huang and H.~Johansson, ``{Equivalent D=3 Supergravity Amplitudes from
  Double Copies of Three-Algebra and Two-Algebra Gauge Theories},''
  \href{http://dx.doi.org/10.1103/PhysRevLett.110.171601}{{\em Phys.Rev.Lett.}
  {\bfseries 110} (2013) 171601},
\href{http://arxiv.org/abs/1210.2255}{{\ttfamily arXiv:1210.2255 [hep-th]}}.

\bibitem{Cachazo:2013iea}
F.~Cachazo, S.~He, and E.~Y. Yuan, ``{Scattering of Massless Particles:
  Scalars, Gluons and Gravitons},''
  \href{http://dx.doi.org/10.1007/JHEP07(2014)033}{{\em JHEP} {\bfseries 1407}
  (2014) 033},
\href{http://arxiv.org/abs/1309.0885}{{\ttfamily arXiv:1309.0885 [hep-th]}}.

\bibitem{Dolan:2013isa}
L.~Dolan and P.~Goddard, ``{Proof of the Formula of Cachazo, He and Yuan for
  Yang-Mills Tree Amplitudes in Arbitrary Dimension},''
  \href{http://dx.doi.org/10.1007/JHEP05(2014)010}{{\em JHEP} {\bfseries 1405}
  (2014) 010},
\href{http://arxiv.org/abs/1311.5200}{{\ttfamily arXiv:1311.5200 [hep-th]}}.

\bibitem{Chiodaroli:2014xia}
M.~Chiodaroli, M.~Gunaydin, H.~Johansson, and R.~Roiban, ``{Scattering
  amplitudes in N=2 Maxwell-Einstein and Yang-Mills/Einstein supergravity},''
\href{http://arxiv.org/abs/1408.0764}{{\ttfamily arXiv:1408.0764 [hep-th]}}.

\bibitem{Antoniadis:1992sa}
I.~Antoniadis, E.~Gava, and K.~Narain, ``{Moduli corrections to gravitational
  couplings from string loops},''
  \href{http://dx.doi.org/10.1016/0370-2693(92)90009-S}{{\em Phys.Lett.}
  {\bfseries B283} (1992) 209--212},
\href{http://arxiv.org/abs/hep-th/9203071}{{\ttfamily arXiv:hep-th/9203071
  [hep-th]}}.

\bibitem{Sen:1995ff}
A.~Sen and C.~Vafa, ``{Dual pairs of type II string compactification},''
  \href{http://dx.doi.org/10.1016/0550-3213(95)00498-H}{{\em Nucl. Phys.}
  {\bfseries B455} (1995) 165--187},
\href{http://arxiv.org/abs/hep-th/9508064}{{\ttfamily arXiv:hep-th/9508064}}.

\bibitem{Borsten:2013bp}
L.~Borsten, M.~Duff, L.~Hughes, and S.~Nagy, ``{A magic square from Yang-Mills
  squared},'' \href{http://dx.doi.org/10.1103/PhysRevLett.112.131601}{{\em
  Phys.Rev.Lett.} {\bfseries 112} (2014) 131601},
\href{http://arxiv.org/abs/1301.4176}{{\ttfamily arXiv:1301.4176 [hep-th]}}.

\bibitem{Anastasiou:2013hba}
A.~Anastasiou, L.~Borsten, M.~Duff, L.~Hughes, and S.~Nagy, ``{A magic pyramid
  of supergravities},'' \href{http://dx.doi.org/10.1007/JHEP04(2014)178}{{\em
  JHEP} {\bfseries 1404} (2014) 178},
\href{http://arxiv.org/abs/1312.6523}{{\ttfamily arXiv:1312.6523 [hep-th]}}.

\bibitem{Anastasiou:2015vba}
A.~Anastasiou, L.~Borsten, M.~J. Hughes, and S.~Nagy, ``{Global symmetries of
  Yang-Mills squared in various dimensions},''
  \href{http://dx.doi.org/10.1007/JHEP01(2016)148}{{\em JHEP} {\bfseries 01}
  (2016) 148},
\href{http://arxiv.org/abs/1502.05359}{{\ttfamily arXiv:1502.05359 [hep-th]}}.

\bibitem{Anastasiou:2014qba}
A.~Anastasiou, L.~Borsten, M.~J. Duff, L.~J. Hughes, and S.~Nagy, ``{Yang-Mills
  origin of gravitational symmetries},''
  \href{http://dx.doi.org/10.1103/PhysRevLett.113.231606}{{\em Phys. Rev.
  Lett.} {\bfseries 113} no.~23, (2014) 231606},
\href{http://arxiv.org/abs/1408.4434}{{\ttfamily arXiv:1408.4434 [hep-th]}}.

\bibitem{Hodges:2011wm}
A.~Hodges, ``{New expressions for gravitational scattering amplitudes},''
  \href{http://dx.doi.org/10.1007/JHEP07(2013)075}{{\em Journal of High Energy
  Physics} {\bfseries 1307} (2013) },
\href{http://arxiv.org/abs/1108.2227}{{\ttfamily arXiv:1108.2227 [hep-th]}}.

\bibitem{Monteiro:2014cda}
R.~Monteiro, D.~O'Connell, and C.~D. White, ``{Black holes and the double
  copy},''
\href{http://arxiv.org/abs/1410.0239}{{\ttfamily arXiv:1410.0239 [hep-th]}}.

\bibitem{Cheung:2009dc}
C.~Cheung and D.~O'Connell, ``{Amplitudes and Spinor-Helicity in Six
  Dimensions},'' \href{http://dx.doi.org/10.1088/1126-6708/2009/07/075}{{\em
  JHEP} {\bfseries 0907} (2009) 075},
\href{http://arxiv.org/abs/0902.0981}{{\ttfamily arXiv:0902.0981 [hep-th]}}.

\bibitem{Bern:2010qa}
Z.~Bern, J.~J. Carrasco, T.~Dennen, Y.-t. Huang, and H.~Ita, ``{Generalized
  Unitarity and Six-Dimensional Helicity},''
  \href{http://dx.doi.org/10.1103/PhysRevD.83.085022}{{\em Phys.Rev.}
  {\bfseries D83} (2011) 085022},
\href{http://arxiv.org/abs/1010.0494}{{\ttfamily arXiv:1010.0494 [hep-th]}}.

\bibitem{Bidder:2005ri}
S.~J. Bidder, N.~Bjerrum-Bohr, D.~C. Dunbar, and W.~B. Perkins, ``{One-loop
  gluon scattering amplitudes in theories with N \&lt; 4 supersymmetries},''
  \href{http://dx.doi.org/10.1016/j.physletb.2005.02.045}{{\em Phys.Lett.}
  {\bfseries B612} (2005) 75--88},
\href{http://arxiv.org/abs/hep-th/0502028}{{\ttfamily arXiv:hep-th/0502028
  [hep-th]}}.

\bibitem{Elvang:2013cua}
H.~Elvang and Y.-t. Huang, ``{Scattering Amplitudes},''
\href{http://arxiv.org/abs/1308.1697}{{\ttfamily arXiv:1308.1697 [hep-th]}}.

\bibitem{Chiodaroli:2015wal}
M.~Chiodaroli, M.~Gunaydin, H.~Johansson, and R.~Roiban, ``{Complete
  construction of magical, symmetric and homogeneous N=2 supergravities as
  double copies of gauge theories},''
\href{http://arxiv.org/abs/1512.09130}{{\ttfamily arXiv:1512.09130 [hep-th]}}.

\bibitem{Bhardwaj:2013qia}
L.~Bhardwaj and Y.~Tachikawa, ``{Classification of 4d N=2 gauge theories},''
  \href{http://dx.doi.org/10.1007/JHEP12(2013)100}{{\em JHEP} {\bfseries 12}
  (2013) 100},
\href{http://arxiv.org/abs/1309.5160}{{\ttfamily arXiv:1309.5160 [hep-th]}}.

\bibitem{Huang:2010rn}
Y.-t. Huang and A.~E. Lipstein, ``{Amplitudes of 3D and 6D Maximal
  Superconformal Theories in Supertwistor Space},''
  \href{http://dx.doi.org/10.1007/JHEP10(2010)007}{{\em JHEP} {\bfseries 1010}
  (2010) 007},
\href{http://arxiv.org/abs/1004.4735}{{\ttfamily arXiv:1004.4735 [hep-th]}}.

\bibitem{Schreiber:2016sss}
A.~Schreiber, ``{Chiral squaring and KLT relations},''
\href{http://arxiv.org/abs/1601.03028}{{\ttfamily arXiv:1601.03028 [hep-th]}}.

\bibitem{Calkins:2014exa}
M.~Calkins, D.~Gates, S.~J. Gates, and B.~McPeak, ``{Is it possible to embed a
  $4D, \mathcal{N}= 4$ supersymmetric vector multiplet within a completely
  off-shell adinkra hologram?},''
  \href{http://dx.doi.org/10.1007/JHEP05(2014)057}{{\em JHEP} {\bfseries 1405}
  (2014) 057},
\href{http://arxiv.org/abs/1402.5765}{{\ttfamily arXiv:1402.5765 [hep-th]}}.

\end{thebibliography}\endgroup

\end{document}